\def\year{2021}\relax
\documentclass[letterpaper]{article} 
\usepackage{aaai21}  
\usepackage{times}  
\usepackage{helvet} 
\usepackage{courier}  
\usepackage[hyphens]{url}  
\usepackage{graphicx} 
\usepackage{natbib}  
\usepackage{caption} 
\title{The Impact of Visualizing Design Gradients for Human Designers}
\author{Matthew Guzdial, Nathan Sturtevant, and Carolyn Yang}\affiliations{

Department of Computing Science, Alberta Machine Intelligence Institute (AMII)\\
    University of Alberta, Canada\\
    \{guzdial, nathanst, carolyn4\}@ualberta.ca}
\begin{document}

\maketitle

\begin{abstract}
Mixed-initiative Procedural Content Generation (PCG) refers to tools or systems in which a human designer works with an algorithm to produce game content. 
This area of research remains relatively under-explored, with the majority of mixed-initiative PCG level design systems using a common set of search-based PCG algorithms. 
In this paper, we introduce a mixed-initiative tool employing Exhaustive PCG (EPCG) for puzzle level design to further explore mixed-initiative PCG. 
We run an online human subject study in which individuals use the tool with an EPCG component turned on or off. 
Our analysis of the results demonstrates that, although a majority of users did not prefer the tool, it made the level design process significantly easier, and that the tool impacted the subjects' design process.
This paper describes the study results and draws lessons for mixed-initiative PCG tool design.
\end{abstract}

\section{Introduction}
\label{sec:intro}

Procedural content generation (PCG) refers to the practice of generating video game content algorithmically \cite{hendrikx2013procedural}. 
When PCG is included in software tools to aid designers or developers it is referred to as mixed-initiative PCG, co-creative PCG, or creativity support tools \cite{liapis2016mixed}.
While mixed-initiative PCG has received increased interest, it is still an under-explored area of research, and there isn't yet a broad literature containing formal human subject study results. 
Further, the majority of mixed-initiative PCG level design systems have relied on genetic algorithms, a type of search-based PCG (SBPCG).
Additional research is required to determine how different types of PCG best fit with different mixed-initiative PCG systems.

One relatively novel branch of PCG is exhaustive or semi-exhaustive PCG (EPCG) \cite{sturtevant2018exhaustive}. 
EPCG employs search, but not the heuristic search of typical SBPCG methods. 
Instead, EPCG exhaustively searches over all possible content, or in the semi-exhaustive case, over some subset of possible content. When EPCG is computationally feasible, the exhaustive search changes the PCG question from ``how to {\em generate} content'' to ``what content to {\em select},'' a choice that can be naturally integrated into a mixed-initiative system.

A recent mixed-initiative tool using EPCG was limited to the autonomous application of a single change \cite{sturtevant2020anhinga}. This was an opaque process that did not involve a designer in the selection process. We propose to instead make the EPCG gradients available to designers, giving them complete control over the selection process.
Further, by studying mixed-initiative EPCG we can address the broader question of how different types of PCG best fit with different mixed-initiative PCG systems.

This paper includes the results of a human subject study on a mixed-initiative EPCG tool for a clone of the game Snakebird called Anhinga.  
Snakebird is a mobile puzzle game, which has appeared in prior EPCG work \cite{sturtevant2020unexpected}.
We built upon an existing open-source implementation \cite{sturtevant2020anhinga},
designing a tool that can visualize the impact of any design change to a user. 
The study is designed to investigate the application and evaluation of mixed-initiative EPCG systems, while broadening the design space of mixed-initiative PCG systems. 

The work presented here has the following contributions:
    (1) A novel EPCG mixed-initiative editor.
    (2) Results of a human subject study, and the analysis thereof.
    (3) Evidence that may suggest that our EPCG editor decreases tedium and eases the designer experience.
    (4) The first evidence of ``deceptive'' mixed-initiative PCG. We present evidence that our editor helps users produce more challenging and complex content, even as users expected the opposite. 


\section{Background}
\label{sec:background}

We begin by looking at related work in PCG and the snakebird domain.

\subsection{Mixed-initiative PCG}

Mixed-initiative PCG refers to systems in which a human works with a PCG approach to produce content \cite{liapis2016mixed,deterding2017mixed}.
Different types of PCG have been included in mixed-initiative systems including constructive PCG \cite{speedtree}, constraint-based PCG \cite{smith2010tanagra}, PCG via machine learning (PCGML) \cite{guzdial2018co,pcgml}, and search-based PCG (SBPCG) \cite{togelius2016search,deterding2017mixed}.
Of these, SBPCG has been the most popular, often employing genetic algorithms or evolutionary search as the PCG component \cite{liapis2016mixed}.
In our work, we employ an under-explored SBPCG method: exhaustive PCG (EPCG), which we discuss in further detail below. 
To the best of our knowledge, this represents the first example of evaluating EPCG in a mixed-initiative context.
 
Mixed-initiative PCG systems have traditionally focused on a small subset of game content, with the majority of existing systems designed to produce platformer game levels \cite{smith2010tanagra,guzdial2018co}. 
There has been relatively less work on mixed-initiative PCG systems for puzzle game design \cite{butler2013mixed}.
Further, the majority of this work has been from the independent and hobbyist community and lacks published analysis.\footnote{\url{https://www.f-puzzles.com/}}\footnote{\url{http://forum.enjoysudoku.com/yzf-sudoku-t36846.html#p284180}}
There is an Angry Birds mixed-initiative SBPCG level design tool \cite{campos2017mixed}. 
Further, Charity et al. \cite{charity2020baba} employed a SBPCG approach for mixed-initiative design of Baba is You levels. 
However, there is still a relative lack of mixed-initiative academic work for puzzle game design compared to platformer game design.
Other mixed-initiative approaches that produce variations on a designer’s current level for non-puzzle games exist  \cite{liapis2013sentient}.
This work is similar to our own, focused on modifying levels rather than producing new levels. However, EPCG allows us to make specific, minimum changes that would not be possible with other SBPCG approaches.

\begin{figure*}[tb]
\centering
\includegraphics[height=0.9\columnwidth]{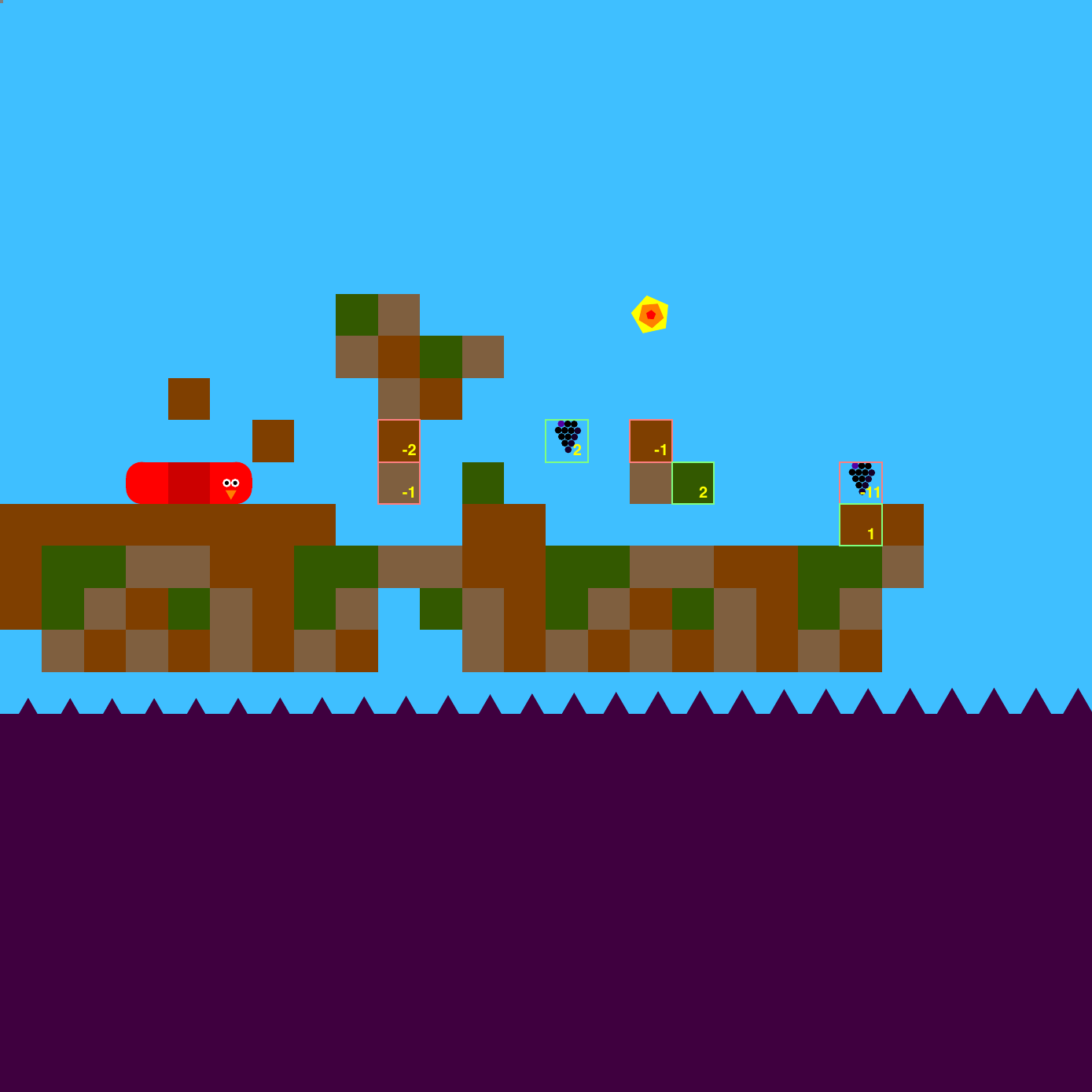}\hspace*{-3pt}
\includegraphics[height=0.9\columnwidth]{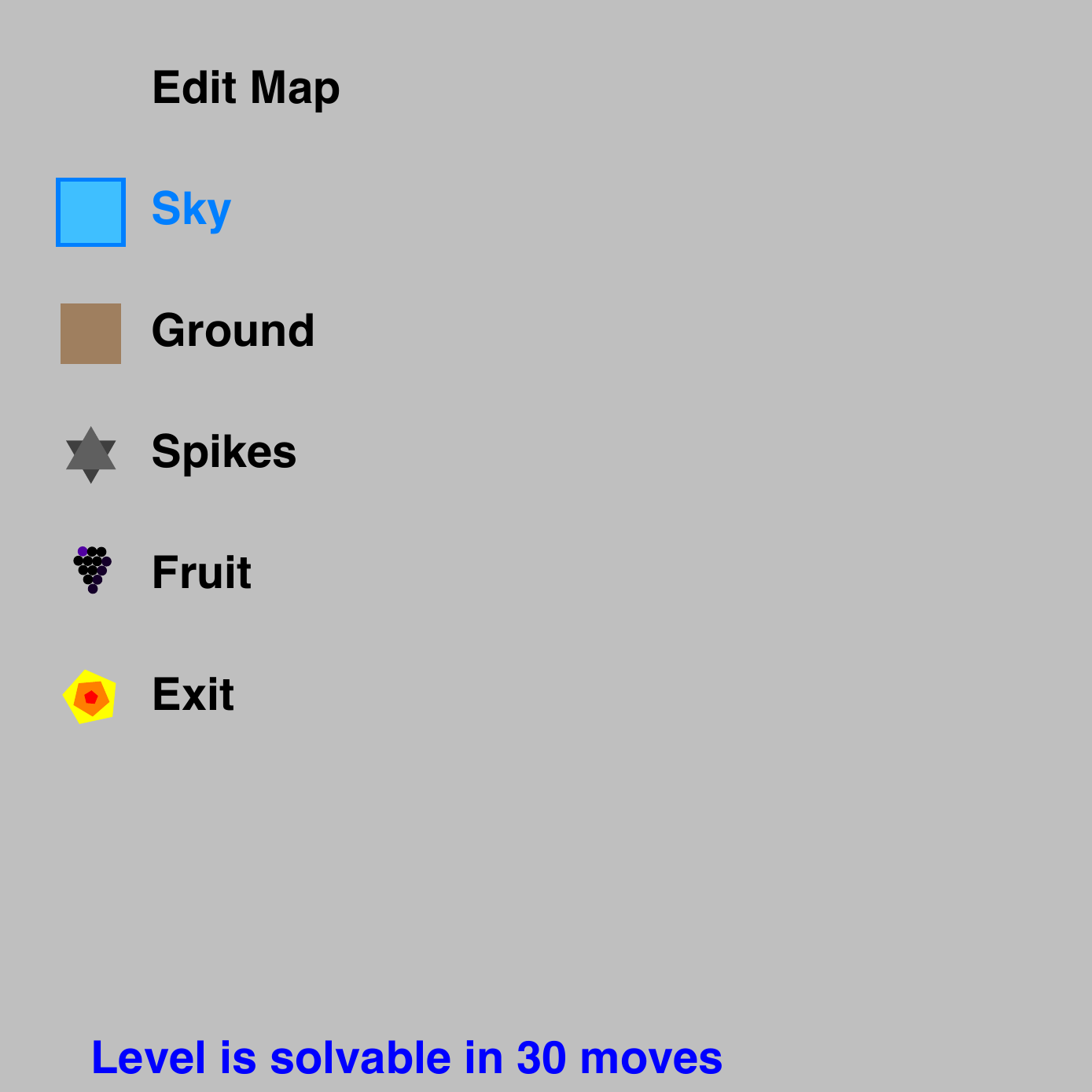}\\
\includegraphics[width=1.8\columnwidth]{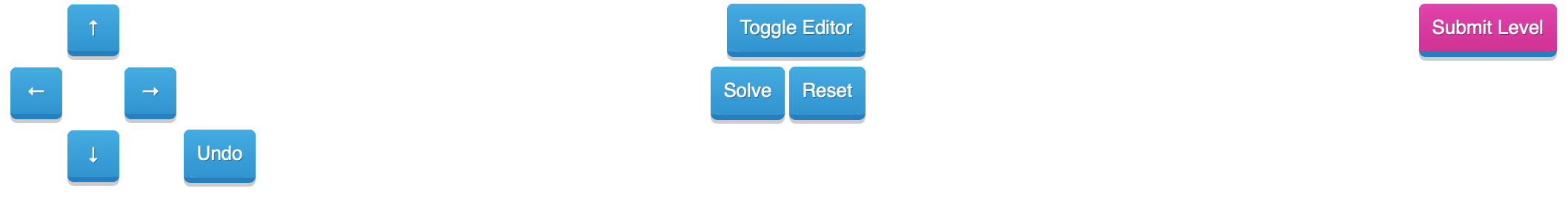}
\caption{Screenshot of the Anhinga Level Editor. The green and red outlined tiles indicate a change in the minimum solution length if that tile is replaced with the currently selected ``Sky''.}
\label{fig:editor}
\end{figure*}

\subsection{Exhaustive PCG (EPCG)}

This work relies on Exhaustive Procedural Content Generation (EPCG), an approach that uses a generator $G$ to exhaustively build content from which an evaluator $E$ selects the best content.
This approach was designed for games \cite{sturtevant2018exhaustive}, but it has also been applied to other fields \cite{gil2020exhaustive}. 
In this work we use EPCG incrementally to create modifications of full levels in Snakebird. 
This implementation is described in detail in prior work \cite{sturtevant2020unexpected}.
At a high level, our EPCG approach identifies the maximum change in solution length from adding or removing a particular type of game object at every location for a puzzle level.

\subsection{Snakebird Domain}

Snakebird is a 2015 game by Noumenon Games, with a 2019 follow-up called Snakebird Primer. 
The goal of each level is to have one or more snakebirds eat all available fruit and then for all snakebirds to leave via the exit.
By eating a fruit, a snakebird grows in length like in the original Blockade (1976), better known as Snake. 
The levels focus on finding the right order to eat the fruit to avoid getting stuck and hazards like spikes and falling.
For more detail, please see the original Anhinga paper~\cite{sturtevant2020unexpected}.

\section{Study Methodology}
\label{sec:methodology}

Our goal for this study was to evaluate the impacts of Exhaustive Procedural Content Generation (EPCG) on mixed-intiative puzzle level design for Anhinga. 
We cover our implementation of an automated, web-based solver, the puzzle level editor interface, and our study design.

\subsection{Solver}

For this work we implemented an automated, web-based solver for Anhinga levels based on the solver originally discussed in \cite{sturtevant2020unexpected}.
We indicate web-based here as we intended our human subject study to run online, which meant that the solver would need to run in the browser. 
We could not simply solve all levels and cache the results since we focus on puzzle level design.
The solver uses a breadth-first search (BFS) to explore a level and return the optimal solution length. 
The major difference between this implementation and the original one was that we bounded the solution search to 50,000 node expansions. 
We did this because the web compiler uses a fixed-memory allocation that cannot grow at runtime.
In EPCG, this solver can be understood as the evaluator $E$, which is half of the EPCG formulation.
We discuss how we employ the results of the solver for our EPCG visualization in the next section.

\subsection{Interface and EPCG Visualizations}

We include a screenshot of the level editor used for the study in Figure \ref{fig:editor}.
On the left-hand side of the editor we have the current level. 
The purple roughly diamond-shaped objects are the fruit the snakebird (in red) must eat to enter the portal or ``exit''. 
This level is a recreation of level 1 from the original Snakebird. 
By using the arrow keys or the buttons on the bottom left side of the screen the user can attempt to play the level, undoing moves with the ``Undo'' button or by pressing the ``q'' key. The user can see the snakebird solve the current level (if possible with our 50K node-limited BFS) by pressing the ``Solve'' button or ``n'' key, and reset the snakebird to its initial position with the ``Reset'' button or ``r'' key. 
On the right-hand side of the screen is the palette of game objects the user can use to edit the level. 
By clicking on a cell with one of these objects selected the user can replace the cell's current object with the selected object. 
If the clicked and selected object match, it will instead be replaced by a sky object. 
The bottom right text ``Level is solvable in...'' updates dynamically with each change as we rerun the solver.

For the EPCG visualization, for every location in the map, we create a level variation as if the user had clicked on the current cell with the current selected object. 
This is equivalent to every single change (exhaustive) that can be made with the currently selected object, which can be understood as the generator $G$ portion of this EPCG implementation. 
These variations are generated
after every change to the level, when a new object is selected, or when a level is loaded. 
We process one cell per frame (the graphics run at 30fps).
The calculations start from the row where the snakebird is located, as that row is more likely to contain interesting changes than the top or bottom row of the map. 
They then continue to successive rows above and below the snakebird until all changes have been analyzed.
For each new, potential level, we compute the solution length with our 50K node-limited BFS (the $E$ portion of the EPCG implementation).
If the modified level was equivalent to the current level in terms of solution length, we do nothing.
If it was distinct from the current level, we calculate the difference in optimal solution length with our solver. 
If this optimal solution is different from the current level then we visualize this gradient with a number indicating the value of the difference in that cell as can be seen in Figure \ref{fig:editor}. 

We give cells with a non-zero difference in values an outline: green if positive and red if negative. 
The shade of green and red ([0.5, 1.0, 0.5] and [1.0, 0.5, 0.5] respectively) were chosen to appear distinct to colorblind participants.
Thus, the ``-1'' in the cell directly ahead of the snakebird in Figure \ref{fig:editor} indicates that replacing the ``ground'' tile at that location with ``sky'' would reduce the optimal solution length by one.
We chose to always have these visualizations on as opposed to require the user to press a button to see them in order to better investigate how they impacted the level design process, as this ensured all participants would see the visualizations. 
This stands in contrast to prior mixed-initiative systems that require a user to ask for AI input \cite{guzdial2019friend}.

\subsection{Study Design}

In this subsection we discuss our human subject study design. 
We included two versions of our editor in our study.
In one version, participants would design a level with the editor setup described in the above section. 
In the second, participants would design a level with the EPCG visualization turned off. 
Notably, even in the second version, users still had the ability to play out the solution with the ``Solve'' button. 
Thus, we can consider the first version to be the full EPCG editor ($E$+$G$) and the second version to to be a half EPCG editor ($E$). 
We chose to include the solver even in the second version as participants had access to it during a series of training levels (discussed below), and we anticipated that participants would strongly prefer to have the solver when designing as well. 
Participants interacted with both versions of the editor. 
Thus, two conditions were the two orders participants could interact with the editors. 

In order to avoid repetition, given that we planned to ask participants to design two levels, we made use of two different ``starting points'' for the two level design tasks. 
We picked two levels (Snakebird 1 and Snakebird Primer 4) that demonstrated initial states with at least one EPCG visualization in our editor. 
We wanted levels that were relatively simple in order to ensure that participants could alter them as they liked. 
We chose to give initial levels as starting points in order to avoid the blank canvas problem \cite{compton2015casual}.
Thus, we ended up with four conditions: the order in which participants saw the two editor versions and the order of the two starting point levels. 

\subsection{Study Process}

Participants were directed to the study through a link posted on social media. 
We posted links on Twitter, Facebook, Slack, and Discord, with the Twitter post receiving the most attention (629 engagements and 15,295 views according to Twitter Analytics). 
From there, participants were brought to an information letter and consent form, which summarized the study and the ethics approval.
At this point, the participants were placed into one of our four conditions by the server. 
The server assigned these conditions sequentially in order to ensure an even distribution of conditions across all who started the study. 
We note that we still had an imbalance in our final results as individuals were more likely to quit the study partway when placed into certain conditions, something we will discuss in Section \ref{sec:results}.

\begin{table*}[tb]
\centering
\small
\begin{tabular}{p{2in}p{2in}p{2in}}
\includegraphics[width=2in]{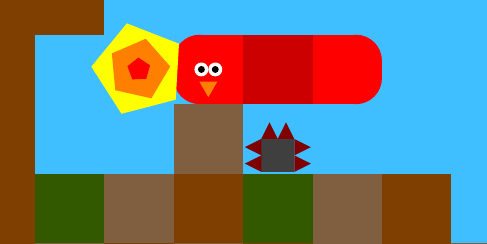} & \includegraphics[width=2in]{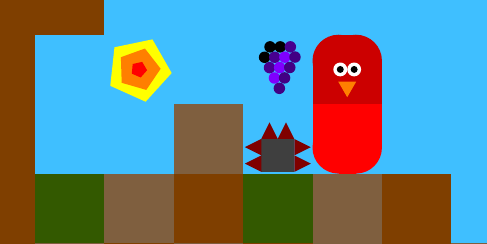} & \includegraphics[width=2in]{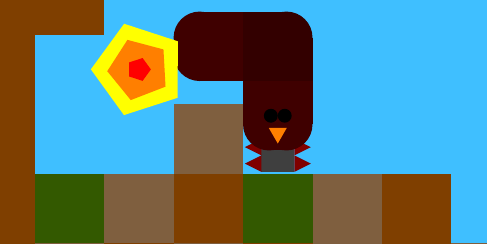} \\
\textbf{1.} The goal of the game is to get your anhinga into the exit & \textbf{2.} Before you can exit, you must eat all the fruit in the level. Eating fruit makes your anhinga longer. \textit{(Note: your anhinga can stand on top of fruit!)} & \textbf{3.} Be careful not to fall on spikes, as this will kill your anhinga. But, if you do, the game will undo your move so you can try again. \\
         
\end{tabular}
\caption{Recreation of the instruction animations and their associated text.}
\label{table:Instructions}
\end{table*}

Participants then came to the main study page. 
At the top of this page were some additional instructions and three animated gifs depicting how to play the game (Table \ref{table:Instructions}). 
From there, participants were asked to play three training levels sequentially (Snakebird Primer 20, Snakebird Primer 27, and Snakebird Primer 23). 
These levels were selected as they demonstrated key mechanics for solving Snakebird levels: standing on fruit, controlling the way your snakebird falls, and avoiding spikes. 
This ensured all participants had at least some experience with snakebird mechanics.

After playing or clicking through the three training levels, participants were asked to complete two level design tasks. 
Based on their condition, participants received a distinct order and mix of the two level editor versions and the two starting levels.
During the design tasks, we logged every participant action and the time for later comparison purposes.

Participants were then asked to fill out an eighteen question survey. 
The first ten questions were ranking questions, asking participants to rank one of their two level design experiences and the final levels according to a number of experiential features. 
We chose ranking questions to avoid the indecision/neutrality problem \cite{joshi2015likert}.
The first three questions took the form ``Which level design experience was more \textbf{blank}?'' with the final word in bold and replaced by ``fun'', ``frustrating'', and ``easier''.
The fourth question was ``Of the two level design experiences, which did you prefer overall?''. 
We chose these experiential features as they match prior work \cite{guzdial2019friend}, and as we anticipated that participants would find it easier to design levels with the EPCG visualizations.

The survey continued with five questions that took the form ``Of the two levels you designed, which do you think players will find more \textbf{blank}?'' with the final word in bold and replaced by ``fun'', ``frustrating'', ``surprising'', ``challenging'', and ``interesting''. 
These matched the experiential features included in the original Anhinga study \cite{sturtevant2020unexpected}, as we hoped to identify any trends in the designer's expectations for future player experiences. 
This was followed by a free text question ``Do you have anything you'd like to share about the level design experiences?''.

The survey ended with seven demographic questions. ``What is your gender?'' allowed for free text responses.
``What is your age?'' had the options: (1) 18-24, (2) 25-30, (3) 31-40, (4) 41-50, and (5) 50+. 
This was followed by two questions around design experience: ``Please pick the category of \textbf{blank} that best fits you:'', with the blank filled with ``puzzle design experience'' and ````Please pick the category of \textbf{puzzle design experience} that best fits you:''. 
These questions had the options: (1) ``No experience'', (2) ``I have designed at least one [blank] before'', (3) I have designed many [blanks], and (4) I am a [blank] design expert. 
These two questions were then followed by two questions around game playing: ``How often do you play \textbf{blank}?'', which was filled by ``puzzle games'' and ``games general'', with answers ``Daily'', ``Weekly'', ``Monthly'', ``Less than monthly'', and ``Not at all''. 
The final question was ``Have you played Snakebird before this study?''.
It had the options ``Yes, I have played the original Snakebird or Snakebird Primer'', ``Yes but only for a previous version of this study'', and ``No''. 
After this final question, participants pressed a button to end the study. 
Due to our ethics agreement, we only include data from individuals who completed the survey.

\section{Study Results}
\label{sec:results}

We had 70 participants start this study.
Of these, only 43 completed the survey. 
We removed three participants from our results due to issues with the participant causing the editor to crash or reloading it. 
For example, one participant made their first level too complex for the online version of the solver, stating in their free text response ``My first level, I designed something that exceeded the AI's solving cap".
With these 40 responses we ran a two-way ANOVA for our two study variables: (1) which level the participants saw first and (2) whether they had the full EPCG editor (with the path length changes visualized) first.
Only the second variable had any impact on the results according to this test, and so we randomly sampled an equal number of individuals who had the full EPCG editor first and second from these 40 participants.
This led to a final set of 34 participants which we used for our statistical analysis. 
Recall that we assigned participants to each condition equally when they started the study, thus we can conclude that participants were more likely to quit the study early if they had the full EPCG editor first, potentially due to being unwilling to design a second level without the full EPCG editor. 
This represents a survivor bias, which we cannot control for in our analysis. 

\subsection{Demographics}

For the 34 participants we included for statistical analysis 17 identified as men, 8 as women, and the remaining 9 gave a variety of responses to the gender question including ``non-binary''/``nonbinary'' and ``genderqueer''. 
7 of our participants indicated they were in the range of 18-24, and the remaining 27 fell in the range of 25-30.
In terms of experience designing puzzles, there was a clear bias towards more experience among our participants. 
Only 11 indicated no puzzle design experience, with 21 indicating at least some puzzle experience, and a further 2 identifying as puzzle design experts. 
This paralleled the results with game design, with the only difference being that three of the respondents with no puzzle design experience indicated they had some game design experience. 
All participants indicated they played games at least monthly, with all but 9 playing puzzle games at least weekly. 
We anticipate the large bias towards puzzle design knowledge came from selection bias, that those with puzzle design experience were more likely to take part, and due to where the study was advertised (followers of games academics are likely to be interested in games). 
However, only two of the participants had played Snakebird before, and that was in a prior study. 
We identify places where this lack of experience may have impacted the results below. 

\begin{table}[tb]
\centering
\begin{tabular}{|l|c|c|c|c|}
\hline
         & Fun & Frustrate & Easy & Prefer  \\
         \hline
Full EPCG     & 12  & 15        & \textbf{27}   & \textbf{6}        \\
\hline
Half EPCG & 22  & 19        & \textbf{7}     & \textbf{28}       \\ 
\hline
\end{tabular}
\caption{Total number of first rankings for each design experience feature. Results in bold represent a significant difference $p<0.005$ using the Wilcoxon Mann Whitney U-test.}
\label{table:DesignPreference}
\end{table}

\subsection{Ranking Results}

\begin{table*}[tbh]
\centering
\begin{tabular}{|l|c|c|c|c|c|c|}
\hline
         & Fun & Frustrate & Surprise & Challenge & Interesting & Prefer  \\
         \hline
Full EPCG Level    & 14  & 15 & 16 & \textbf{9} & 15 & 16 \\
\hline
Half EPCG Level & 19  & 18 & 17 & \textbf{24} & 18 & 17 \\
\hline
\end{tabular}
\caption{Total number of first rankings for each expected player experience feature. Results in bold represent a significant difference $p<0.005$ using the Wilcoxon Mann Whitney U-test.}
\label{table:LevelPreference}
\end{table*}

\begin{table*}[tb]
\centering
\begin{tabular}{|l|c|c|c|c|c|c|c|c|}
\hline
         & \multicolumn{2}{c|}{Time (min)} & \multicolumn{2}{c|}{Num. Actions} & \multicolumn{2}{c|}{Num. Solver} & \multicolumn{2}{c|}{Sol. Length}  \\
         \hline
         & mean & median & mean & median & mean & median & mean & median \\
         \hline
Full EPCG     & 61.43$\pm$297.53 & 5.05     & 141.65$\pm$62.08 & 143        & 5.94$\pm$4.44 & 3.5   &  60.41$\pm$48.54 & 49 \\
\hline
Half EPCG & 7.15$\pm$9.54 & 3.84       & 180.79$\pm$147.1 & 163        & 8.5$\pm$8.13 & 6.5     &  45.18$\pm$32.32 & 39 \\ 
\hline
\end{tabular}
\caption{Quantitative results based on logged data of participants during the design process.}
\label{table:QuantitativeResults}
\end{table*}
We used the Wilcoxon Mann Whitney U-test for statistical analysis of the ranking questions results, given that the values are non-numeric and do not follow the normal distribution. 
We ran this test between subjects to determine if the differences in distribution were significant for each experiential feature. We ran the test for a total of ten times, and used $p<0.005$ as a significance threshold according to the Bonferroni correction method. 
Our first set of questions asked participants to rank the two level design experiences, one with the EPCG visualizations turned on (Full EPCG) and one with the EPCG visualizations turned off (Half EPCG). 
We summarize these results in Table \ref{table:DesignPreference}.
In each cell we give the number of individuals who ranked that design experience first across these different experiential features.

For fun and frustration we found no significant difference between the two experiences. 
For frustration, we found that 30 of the 34 participants ranked the second design experience as more frustrating, which suggests that participants did not like designing the second level, regardless of condition. 
We neutralize this effect by analyzing the Full vs. Half EPCG results with an equal number of participants in both ordering conditions. 
We found that the participants ranked designing levels with the Full EPCG editor as significantly easier.
This matches our expectations, that the EPCG visualizations would help to guide the design experience. 

The participants indicated that they much preferred the Half EPCG version of our tool without the EPCG visualizations. 
This went against our expectations, but some of the comments clarify why. 
One participant reported ``I found suggestions to be intrusive to my creativity'', which may indicate some participants felt too constrained by the visualization even though they could ignore it. 
Another participant reported ``I found that the tool that uses AI to tell me how many more steps I am including really hampered my ability to come up with a good idea for a puzzle... I [was] constantly worrying about how many steps I was taking away or removing'', which suggests that the visualizations may have been overwhelming or stressful to some participants. 
Notably, individuals could not control when the AI made these ``suggestions'' since they were visualized automatically with every edit. 
While these kind of automated suggestions have been previously explored in the mixed-initiative literature \cite{guzdial2017general}, this is the first instance of a study demonstrating a strong negative reaction to them by users.
On the other hand, both of these individuals expressed appreciation for the solver saying ``I liked... the auto-solver'' and ``I did find the solve function very useful'' respectively. 
Given that our solver is one-half of the EPCG implementation, we do not take this as a refutation of EPCG, but of our implementation for conveying the EPCG results to users. 
Were we to run the study again, we would add a button allowing users to hide/show the design gradients. 

In Table \ref{table:LevelPreference} we give the results of our level ranking questions using the same format as the design ranking results. 
These questions asked participants to rank how a hypothetical player would experience their two levels, and were included in an attempt to capture design intent.
The results in each row sum to 33 instead of 34 as one participant left all these questions blank, which participants could have done for any survey question.
While the most fun level, most surprising, and most preferred level results appear to be close to a random distribution, this is not the case. 
Instead, participants across both conditions ranked their second level as more fun at a ratio of 20:13, their second level as more surprising at a ratio of 21:12, and their second level as more preferred at a ratio of 23:10. 
This indicates a clear bias towards the second level regardless of condition, which we interpret as an indication that participants were learning Snakebird level design as they went. 
As such, and due to the participants lack of experience with Snakebird, they may have felt that the second level was an improvement on their first. 
This stand in comparison to users finding the second design task more frustrating, indicating users felt the output levels were better even if the design experience was not ideal. 

Our only statistically significant result was that participants expected players to find their level produced with the Half EPCG editor (no visualizations) more challenging.
The participants also split their frustration and interestingness rankings nearly evenly.
However, from prior work with EPCG and Anhinga we can expect solution length to have a significant, moderate correlation with these three features of player experience \cite{sturtevant2020unexpected}.
The solution lengths for the levels produced with the EPCG visualization were 50\% longer (46 steps to 66 steps).
Thus, we would expect players to rank these levels as more challenging, frustrating, and interesting in contrast to the designer rankings. 
We suspect that the participants may have conflated the experience of designing the level with the experience of playing the level. 
Participants had a significantly harder time designing levels with the Half EPCG editor according to their rankings, and so they may have expected a hypothetical player to experience the same challenge. 

\subsection{Quantitative Results}

We extract four metrics from the log data of the participants: the amount of time in minutes spent on a design task (Time), the number of actions the participants took (Num. Actions), the number of times the participants used the solver (Num. Solver), and the solution length of the final level (Sol. Length).
We give the mean, standard deviation, and the median of these metrics for the two types of level design experiences in Table \ref{table:QuantitativeResults}.
Participants clearly took more time working with the Full EPCG editor than the Half EPCG editor, though the large mean time is due to a single two-hour outlier. 
The median values are only slightly above 1 minute apart. 
This is in contrast to the self-reported rankings, which indicated that the Full EPCG editor with the visualization was easier to work with and therefore, one might assume, faster.
Part of the issue may have been that the EPCG visualization added significantly more information to analyze, which could account for the time difference. 

Based on Table \ref{table:QuantitativeResults}, our participants performed many more actions and used the solver more times in the Half EPCG editor. 
This is in contrast to the amount of time spent with each editor, and further supports our hypothesis about the visualizations adding extra analysis/processing time for users. 
While we did not include a use case without the solver, we note that of the 34 participants only 2 did not use the solver across both editors, and only one participant used it once.  
Since the solver saw significant use, we take this to indicate its value to participants, which is also supported from participant free text answers including ``The Solve button by itself was quite helpful!'' ``It was fun seeing the tool execute exactly the solution I had in mind, for both levels! :)'' and ``I kind of wish I had the Solve button in my actual puzzle games I make!''. 
While prior work has investigated including automated players in mixed-initiative tools \cite{hoyt2019integrating}, this is the first human subject study to provide evidence for its positive impact. 

The solution lengths of the final levels indicate that the levels produced with the Full EPCG editor were roughly 50\% longer on average, though only about 25\% longer in the median case. 
This suggests that the EPCG visualizations did help participants make levels with longer path lengths, though the difference is not significant using statistical tests. 
Still, as we noted above, this provides some evidence that players would find the levels produced with the Full EPCG editor more frustrating, interesting, and challenging \cite{sturtevant2020unexpected}.
We provide further evidence that the Full EPCG editor led to more challenging levels in contrast to designer expectations in a metric-based analysis below.

\subsection{Study Comments}

We lack the resources to fully analyze the comments left by participants to our free response question. 
However, we found that in general they broke into five categories. 
The first two were positive and negative reactions to the tool generally. 
We've already included examples of these two categories throughout this paper, but include two more here for reference. 
On the positive side ``The tool just helped remove some tedium'' and on the negative side ``I feel like I understand my levels less well because I had an autosolver''. 

The next category we identified were suggestions for alternatives to the optimal path as an EPCG evaluator $E$. 
These were some of the longer comments and clearly indicated that the participant had experience with Artificial Intelligence. 
For example ``The info in the [Full EPCG] editor helped me pick spots that would make for more interesting (challenging) puzzles, by finding particular spots that were local maxima of increased solution length. In particular, there were two classes of spot I could identify: those that increased solution length via backtracking/travel, and those that increased it by requiring a different approach. The latter were very valuable, the former less so.''

Our final two categories of comments were complaints about the survey specifically and comments that suggested a misunderstanding of the task. 
The most common complaint, occurring twice, concerned the lack of a neutral option for the ranking question: ``the survey only has options for `first' or `second'; some of my answers would more accurately be `about the same'''. 
The misunderstanding comments included misunderstandings about Snakebird mechanics:``Is being able to use the food tiles as solid tiles intended?''. Alternatively, another common thread was a mistaken belief that we were designing a novel game: ``I think this game could benefit from ai targets with rules. things that move when you do something or follow patterns''. 

\section{Metric Analysis}
\label{sec:metric_analysis}

In this section we describe the results of a metric-based analysis of the study levels. 
Prior to the study, we identified a metric we call \emph{Solution Density}.
We define this metric as the maximum number of times that the Snakebird's head was in the same cell position during the optimal solution. 
We implemented this metric with our automated solver, which can impact the metric. For instance, the solver tie-breaks the actions down, up, left, and right in that order.
Thus, if multiple moves are equally good, our solver will ``prefer'' to move up before left. Based on our experiences with Snakebird, we hypothesized that this metric might correlate to more interesting or challenging levels.
In these levels, a snakebird must be manipulated carefully, which often involves revisiting the same location.

We evaluated this metric by comparing it to the original Snakebird and Snakebird Primer levels.
We selected only those levels that matched the planned constraints of our study, a single snakebird, and none of the more complex game objects (e.g. portals or blocks). 
This led to only 10 levels for Snakebird, and 29 for Snakebird Primer.
Snakebird, which is considered a harder game than Snakebird Primer, has a median Solution Density of 3 for it's levels, compared to Snakebird Primer's 2. 

Each level has a number associated with it, which roughly corresponds to their designer-intended sequence in these games. 
We can also expect this number to correspond with challenge or complexity (i.e. level 1 being the easiest). 
Thus, if our metric correlates with these sequence indices then that supports that the metric corresponds with the Snakebird designer's concept of challenge or complexity. 

We used Spearman's Rho for our correlation test as the Solution Density values did not follow a normal distribution. 
There were insufficient levels that met our constraints in the original Snakebird to find a significant correlation ($rho=0.55$ $p=0.096$). 
However, Snakebird Primer had sufficient levels, leading to a significant and moderate correlation ($rho=0.51$ $p=0.005$). 
Thus, we conclude that our Solution Density is an appropriate measure for analyzing levels in terms of challenge or complexity.

We ran one last statistical test using the paired version of the Wilcoxon Mann Whitney U-test, representing our only within-subjects comparison. 
We wanted to determine whether individuals were more likely to produce levels with higher Solution Density values with the tool than without it. 
Our test indicated this was the case with a significance threshold of $p<0.05$, with the Full EPCG-produced levels having a median Solution Density of 3 compared to the Half EPCG levels' median Solution Density of 2. 

\section{Discussion and Takeaways}
\label{sec:discussion}

In this section we discuss our results in terms of takeaways for mixed-initiative PCG. 
While some did like the Full EPCG visualizations (``I found [the EPCG visualization] incredibly helpful especially in trying to create a surprising to me solution.''), the majority of participants appeared to not prefer our implementation. 
We believe this was due to the visualized ``suggestions'' overwhelming or frustrating users. 
We note that from our own experience, we prefer to design with the suggestions as we ignore them except when we want to use them. 
Thus, we expect that reducing the number of suggestions (only including changes above some magnitude) or only showing them after a user request (i.e. with an ``Analyze'' button) would be preferable to most users. 
However, we intentionally chose for the visualizations to be ``always on'' to better evaluate how they impacted the design task. 

Most participants had more fun when they played the levels themselves (``I had a lot more fun... as I immediately went in to try out the changes I was making''), but participants also responded positively to the solver. 
We expect this was potentially due to participants having agency over when the solver ran, if at all. 
However, participants also seemed to conflate their experience playing their own level with how other players would respond to their level. 
Participants ranked their Half EPCG levels as being more challenging from a hypothetical player's experience, but our metric-analysis and the solution lengths casts doubt on this.
Instead our analysis suggests that the Full EPCG approach did help the participants make more challenging levels, even as they were unaware of this effect. 
We refer to this effect as deceptive mixed-initiative PCG.
We intend to run a follow-up study with participants playing these levels to confirm whether this mismatch of designer expectation and player-perceived challenge holds. 

While most of our participants did not have direct experience with Snakebird, games where a snake eats fruit and grows longer are not unusual.\footnote{\url{https://joelthefox.github.io/2019-08-21-Snake-Puzzle-Games/}} 
Thus, despite this lack of experience, some participants quickly came up with puzzles (``I still used my usual approach of finding an interesting interaction or deductive path, and designing the puzzle around that.''). 
We anticipate that if we employed a more novel or unusual set of game mechanics and dynamics, users may have been more likely to rely on the EPCG components.

\section*{Acknowledgements}

This work was funded by the Canada CIFAR AI Chairs Program. We acknowledge the support of the Alberta Machine Intelligence Institute (Amii). We acknowledge the support of the Natural Sciences and Engineering Research Council of Canada (NSERC).

\bibliographystyle{aaai21.bst}
\bibliography{aiide21.bib}

\end{document}